# Accelerated Discovery of 3D Printing Materials Using Data-Driven Multi-Objective Optimization

Timothy Erps, [1][†] Michael Foshey, [1][†][*] Mina Konaković Luković, [1] Wan Shou, [1][*] Hanns Hagen Goetzke, [2] Herve Dietsch, [2] Klaus Stoll, [2] Bernhard von Vacano, [2] Wojciech Matusik[1]

[1]Computer Science and Artificial Intelligence Laboratory (CSAIL), Electrical Engineering and Computer Science Department, Massachusetts Institute of Technology, Cambridge, MA, USA.

[2]BASF SE, RA/OIS, Carl Bosch Str 38, 67056 Ludwigshafen, Germany.

*Corresponding author. Email: mfoshey@mit.edu, wanshou@mit.edu

†These authors contributed equally to this work.

## Abstract

Additive manufacturing has become one of the forefront technologies in fabrication, enabling new products impossible to manufacture before. Although many materials exist for additive manufacturing, they typically suffer from performance trade-offs preventing them from replacing traditional manufacturing techniques. Current materials are designed with inefficient human-driven intuition-based methods, leaving them short of optimal solutions. We propose a machine learning approach to accelerate the discovery of additive manufacturing materials with optimal trade-offs in mechanical performance. A multi-objective optimization algorithm automatically guides the experimental design by proposing how to mix primary formulations to create better-performing materials. The algorithm is coupled with a semi-autonomous fabrication platform to significantly reduce the number of performed experiments and overall time to solution. Without any prior knowledge of the primary formulations, the proposed methodology autonomously uncovers twelve optimal composite formulations and enlarges the discovered performance space 288 times after only 30 experimental iterations. This methodology could easily be generalized to other material formulation problems and enable completely automated discovery of a wide variety of material designs.

## Introduction

Additive manufacturing is an emerging technique to manufacture objects with complex structures and functions[1,2]. Recently, glass[7], batteries[8-10], high-temperature ceramics[11], and artificial organs[12] have been successfully 3D printed. Among various polymer printing methods, stereolithography (SLA) and material jetting 3D printing have shown promising applications such as robotic assemblies[3], prosthetics[4], biologic scaffolds[5], and customized goods (e.g., footwear)[6]. However, the development of new 3D printing materials currently relies on domain knowledge of polymer chemistry and extensive experimental trials, constraining the efficiency and scalability of materials development. Furthermore, 3D printing materials are typically designed and optimized using one factor at a time (OFAT)[30]. This approach often requires testing an excessive number of samples, generating large waste and undesirable environmental impact, while not always discovering the most optimal solutions. In order to make additive manufacturing a more widely adopted manufacturing approach, it is critical to accelerate the development of materials with optimized performance. Additionally, to address the challenges of diverse application domains, such as bioengineering and aerospace engineering, material performance needs to be optimized for a specific application.

Different autonomous systems have been recently proposed to accelerate material development[12-16] and significantly reduce human labor. Coupled with automation, modern optimization techniques have the ability to simplify the process of optimizing materials for given performances[17]. However, the cost and time per experiment are often high, material supply might be limited, and collecting large amounts of data is impractical. Thus, data-efficient optimization algorithms are preferred. In addition, for many real-world applications, multiple performance criteria should be met. Satisfying multiple objectives simultaneously increases the complexity of performance space search and the ability to discover an optimal solution. In this scenario, a multi-objective optimization approach that guides the sampling of design space can be used to efficiently reduce the number of performed experiments[18].

In this work, we propose a semi-automated data-driven workflow (summarized in Figure 1) for discovering new photocurable inks for additive manufacturing. The semi-automated pipeline is developed to be cost-effective and efficient for discovering 3D printing materials; however, a completely autonomous system is possible with certain robotic manipulators[20,21]. The aim of the workflow is to find a set of best composite formulations composed of six primary formulations of photocurable inks, to improve the mechanical properties beyond the performance levels of primary formulations designed by hand. These composite formulations are automatically optimized for multiple performance objectives with a limited amount of experiments performed.

The proposed workflow starts by dispensing the primary formulations on-demand in a specific ratio (Figure 1a) and then thoroughly mixing them (Figure 1b) to prepare a composite formulation. Each composite formulation is then transferred into a jet-valve 3D printer for sample fabrication (Figure 1c and Figure S2, Supporting Information), followed by post-processing (Figure 1d) to complete the sample preparation. The samples are then tested to extract their multiple quantitative mechanical performance parameters (i.e., toughness, compression modulus, and maximum compression strength) (Figure 1e and Figure S4). To minimize the resources spent on testing different formulations and rapidly discover better performing designs, we utilize a data-driven approach based on Bayesian optimization (Figure 1f). This optimization approach automatically learns from prior experiments to inform future decisions on which formulations to test next. A key insight in decision-making lies in balancing between exploiting the most promising formulations and exploring the uncertain regions of the design space. We iterate through the workflow until the experimental budget is reached.

We demonstrate the rapid performance space improvement and discovery of twelve 3D printing materials with optimal trade-offs after only 30 algorithm iterations. Our method can be easily generalized to other formulation design problems, such as optimization of tough hydrogels[33], surgical sealants[34], or nanocomposite coatings[35].

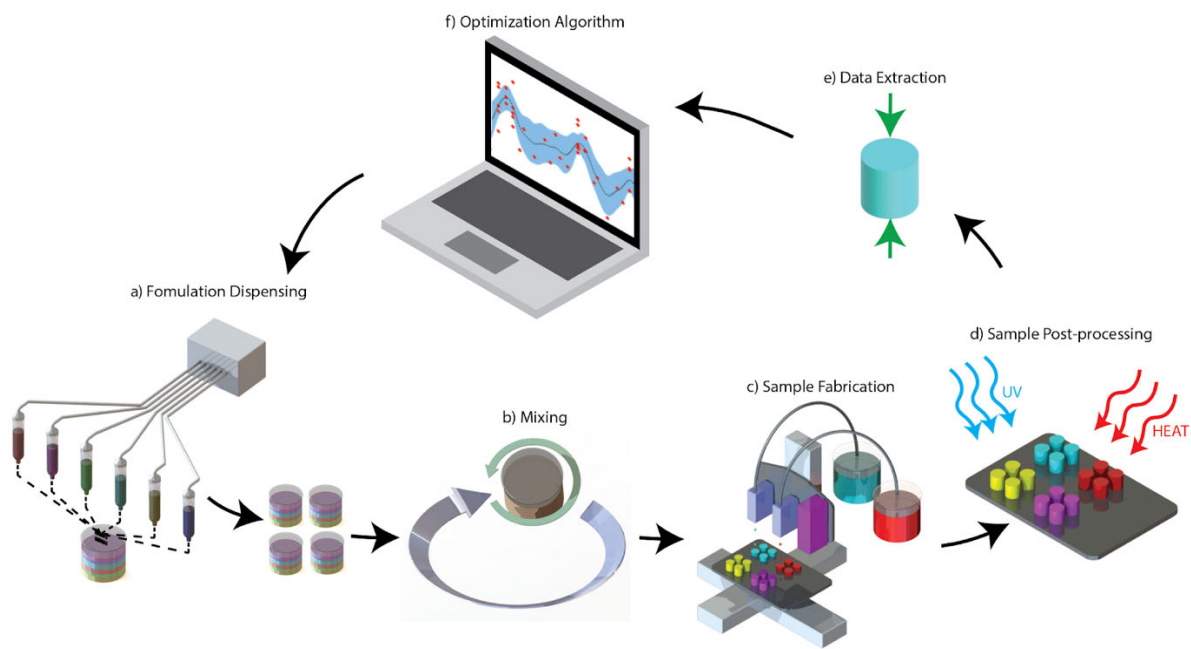

**Fig. 1. Schematic workflow of the accelerated materials discovery system**. a. Primary inks dispenser for dispensing desired formulations; b. Speed mixer for formulation homogenization; c. Jet-valve 3D printer for sample fabrication; d. Sample post-processing with UV curing and heating; e. Compression test for performance data extraction; f. Bayesian optimization algorithm for formulation and performance evaluation, and suggestion for which new formulations to test.

## Results

### Base ingredients and material formulations

We begin by generating a set of photocurable primary formulations that are compatible with each other to mix and have diverse mechanical properties. As key materials for 3D printing, photocurable monomers and oligomers are widely used in light- and ink-based printing[22, 23]. Thus, the development of photopolymers' formulation libraries would be a substantial step to meet customized printing requirements. Instead of developing printing materials from scratch, we start by identifying 8 commercially available formulation ingredients (1 photoinitiator, 3 diluents, and 4 oligomers), as illustrated in **Figure 2** (details can be found in Materials and Methods Section). Then, 6 primary formulations (A, B, C, D, E, and F) are made up of the 8 formulation ingredients in the library. To ensure all possible combinations of formulation ingredients are 3D printable, the primaries are designed to be within a printable regime of viscosities. Surfactant is also added to adjust the surface tension, increasing compatibility with the printer. Photoinitiator is set as a constant across all formulation primaries to avoid its influence on sample performance. Primary formulations are preselected for their ease of printability, print viscosity, and uniqueness of mechanical properties and chemical composition compared to others in the initial screening. However, the selection of materials can be easily expanded depending on the printing method and specific application.

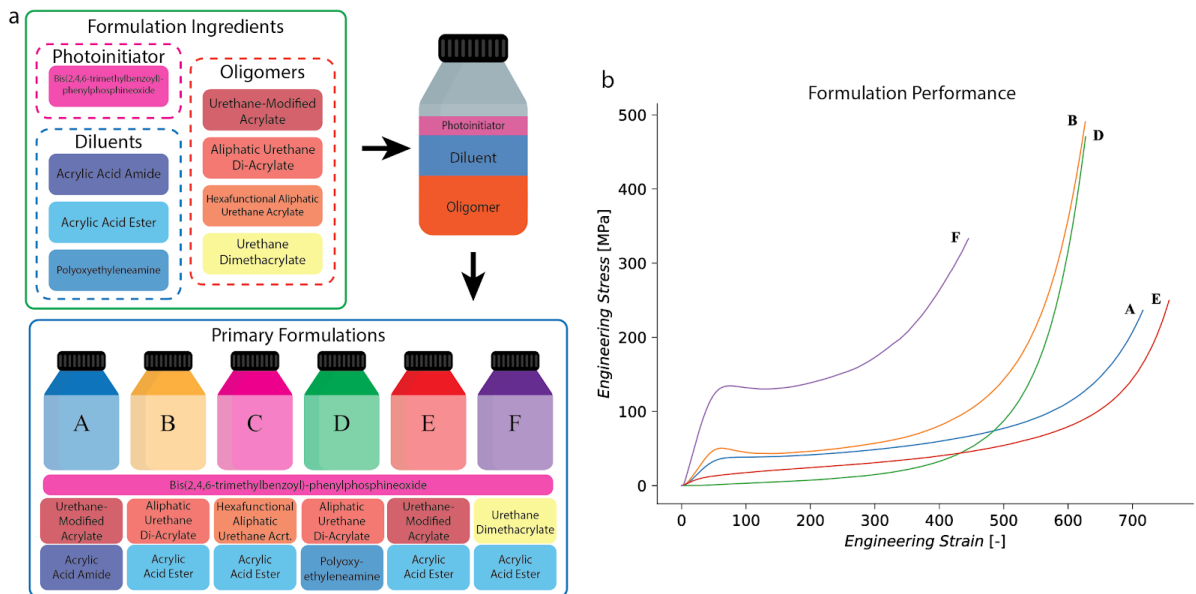

**Fig. 2. Primary formulations A, B, C, D, E, F used in our system.** (a) Primary formulations are made up from 8 formulation ingredients (1 photoinitiator, 3 diluents, and 4 oligomers). (b) Primary formulations performance. The primary formulations are designed to cover a broad range of mechanical properties. Formulation C was too brittle to post-process and test.

A composite formulation is created by mixing the 6 primary formulations in a desired ratio. These primary formulations are automatically dispensed by a 6 channel syringe pump system (details can be found in Materials and Methods Section), and subsequently homogenized in a dual asymmetric centrifuge mixer. These two steps reduce the time and variation for sample preparation, resulting in a set of inks with different formulations that are suggested by the optimization algorithm.

**Jet-valve 3D Printing**

To increase the variety of inks that can be printed and reduce the printing set up time, we use a 3D printing process based on jet-valve dispensing technique. When compared to other types of 3D printing techniques, jet-valve has the ability to dispense inks with a larger variety of fluid properties[24]. Furthermore, jet-valve dispensing requires significantly less tuning of process parameters to attain reliable printing. These traits increased ink variety that could be tested and reduced the time for sample fabrication and data collection. In order to reduce unnecessary variations in geometry and degree of polymerization caused by 3D printing, samples are post-processed via UV curing and heating to ensure complete reaction of components and are flattened via a machining process to eliminate variations in cylinder height (details can be found in Materials and Methods Section).

Due to the layer-by-layer fashion of 3D printing, processing introduces a number of features, such as layer to layer adhesion, that can affect the resulting mechanical properties of the printed material[25]. Hence, testing 3D printed composite formulations ensures that processings affecting mechanical performance are included in the optimization. In an extended optimization scheme, 3D printing parameters can explicitly be included for optimization. Finally, to extract the performance data from each formulation, the 3D printed and post-processed samples are compression tested using a universal tester.

**Multi-objective optimization with batch evaluations**

Composite formulations with high degrees of primary C (consisting of hexafunctional aliphatic urethane acrylate, expectedly yielding the highest possible crosslinking density in cured materials by weight and acrylic acid ester) turned out to be too brittle to flatten for compression testing. Therefore, we imposed a maximum fraction of 50% of C in a formulation, leading to the following constraints on the formulation design: (1) *a, b, d, e, f* ∈ [0,1]; (2) *c* ∈ [0,0.5]; (3) *a+b+c+d+e+f* = 1, where *a, b, c, d, e,* and *f* refer to the amounts of each of 6 primaries, scaled to be between 0 and 1 in the ratio of the sample weight. We further use *a-f* as variables in our optimization system (indicated in **Figure 3**).

The goal of the optimization algorithm is to navigate the 6-dimensional design space of primary formulation ratios *a-f* and quickly uncover the best performing designs with respect to three objectives: toughness, compression modulus, and maximum strength. These performance objectives are chosen because they are mechanical properties that are important for designing and selecting structural materials in engineering applications[27]. Typically, all three of these material properties need to be maximized for many engineering applications. However, these objectives can often be conflicting[19]; hence, there is no single optimal solution, but rather a set of best performing designs with different trade-offs. Depending on the application, the higher performance of one of these properties can be more important than others. The set of such solutions is formally called *Pareto set*, and their corresponding values in the performance space are called *Pareto front*. A point is considered to be *Pareto optimal* if improvements in one objective can only be achieved if at least one other objective value is decreased. The quality of the Pareto front is measured with a *hypervolume indicator*, *i.e.,* a volume of the region of the performance space dominated by the points on the Pareto front (see Figure S5). Here our goal is to find the Pareto front with the largest possible hypervolume indicator.

The design space is a 6-dimensional space with five parameters that can take real values between 0 and 1, and the sixth parameter varies between 0 and 0.5 (due to the limitation imposed on C; a detailed description of design space is available in Materials and Methods Section). Randomly or intuitively sampling this design space may take months or years to gain enough knowledge of the samples' performance and uncover optimal regions of the space. To make the optimization more time- and cost-efficient, we utilize a data-driven approach. The approach learns to predict the performance of untested samples and guides the sampling of the design space to quickly find better performing designs. More specifically, we follow the Bayesian optimization strategy that has proven effective for applications with black-box objective functions and a limited budget of tested samples[26]. Our approach adapts the algorithm proposed by Bradford et al. [18] that solves a multi-objective Bayesian optimization problem, as we aim to optimize three objectives simultaneously. The optimization algorithm consists of four main parts, summarized in **Figure 3**. More details on the algorithm are available in the Materials and Methods Section. To further reduce the experimental time, in each algorithm iteration we evaluate a batch of four different samples in parallel.

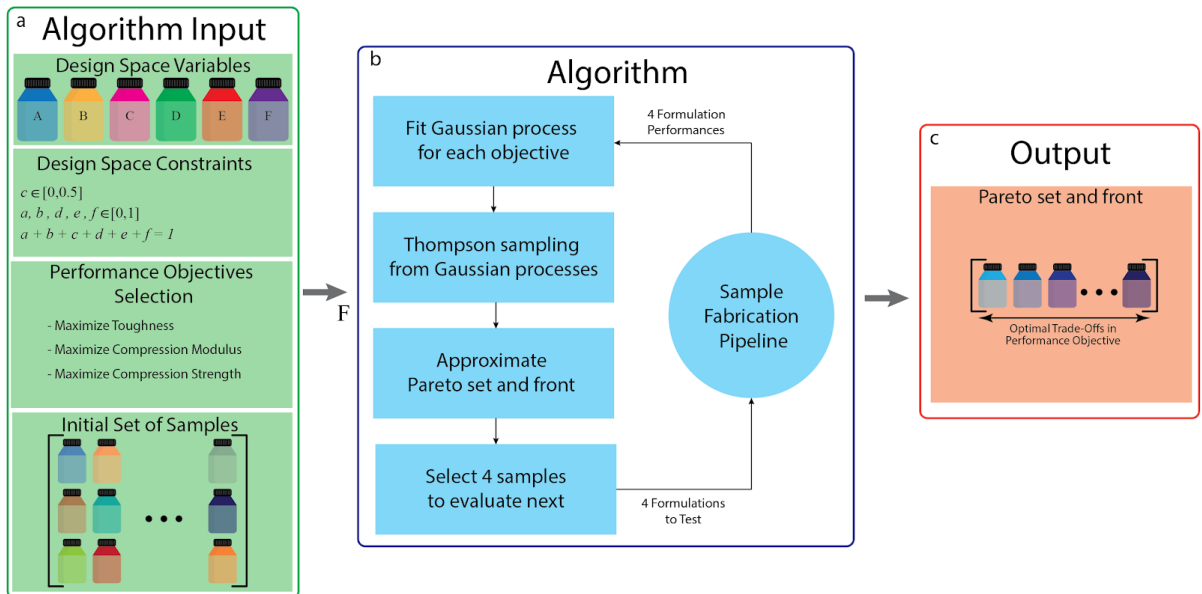

**Fig. 3. Overview of the optimization algorithm used to discover optimal 3D printing material formulations.** (a) The information needed for input include the variables within the design space, design space constraints, performance objectives the algorithm is trying to attain, and an initial set of data that is utilized by the algorithm to build an initial model. (b) An overview of the iterative steps taken by the algorithm to suggest new formulations that may have higher performance. (c) The output from the algorithm after it has finished optimization.

**Discovered materials with optimal trade-offs**

To test the proposed material development workflow, we conduct 30 algorithm iterations in total, as our budget is fixed to 120 samples in addition to the initial dataset. After testing 150 samples (30 initial samples and 120 proposed by the algorithm), the system was able to identify a set of 12 formulations that had optimal trade-offs in the 3 mechanical properties of compression modulus, maximum compression strength, and toughness, as shown in **Figure 4**. The Pareto set includes formulations that increase the maximum compression strength by 70.8% and toughness by 50.8% over the performance of the initial 6 primaries.

The optimization found the final points on the Pareto set at iterations 0 (as the primary formulation F is Pareto optimal), 10, 18, 20 (2 points), 21, 23, 24, 25, 27, 28, and 30. The formulations with the highest performing compression strength and toughness are found at iteration 28 and 23, respectively. Furthermore, the hypervolume did not attain a steady-state value before the budget was reached. By further optimizing, more formulations could be attained. After 30 iterations, the optimization algorithm increased the hypervolume indicator of the performance space by a factor of 1.65 while our pipeline gave an overall improvement by a factor of 3.25 (see Figure 4b). The hypervolume of the formulation primaries, initial data set, and final Pareto front are 2.86e+07, 5.64e+07, and 9.32e+07, respectively. This improvement means that by further optimizing the current set of formulation primaries, a much broader set of performance parameters can be attained. The formulations in the Pareto set vary in compression strength from 308 MPa to 697 MPa. The highest compression strength of the initial 6 primary formulations was 435 MPa of primary B. Compression modulus of optimal solutions spans from 1.1 MPa to 2.93 MPa. The optimization yielded no composite formulation with a higher compression modulus than primary F; hence, the pure primary F formulation lies on the final Pareto set. Pareto set toughness varies from 68.6 GPa of pure primary F to 103.6 GPa.

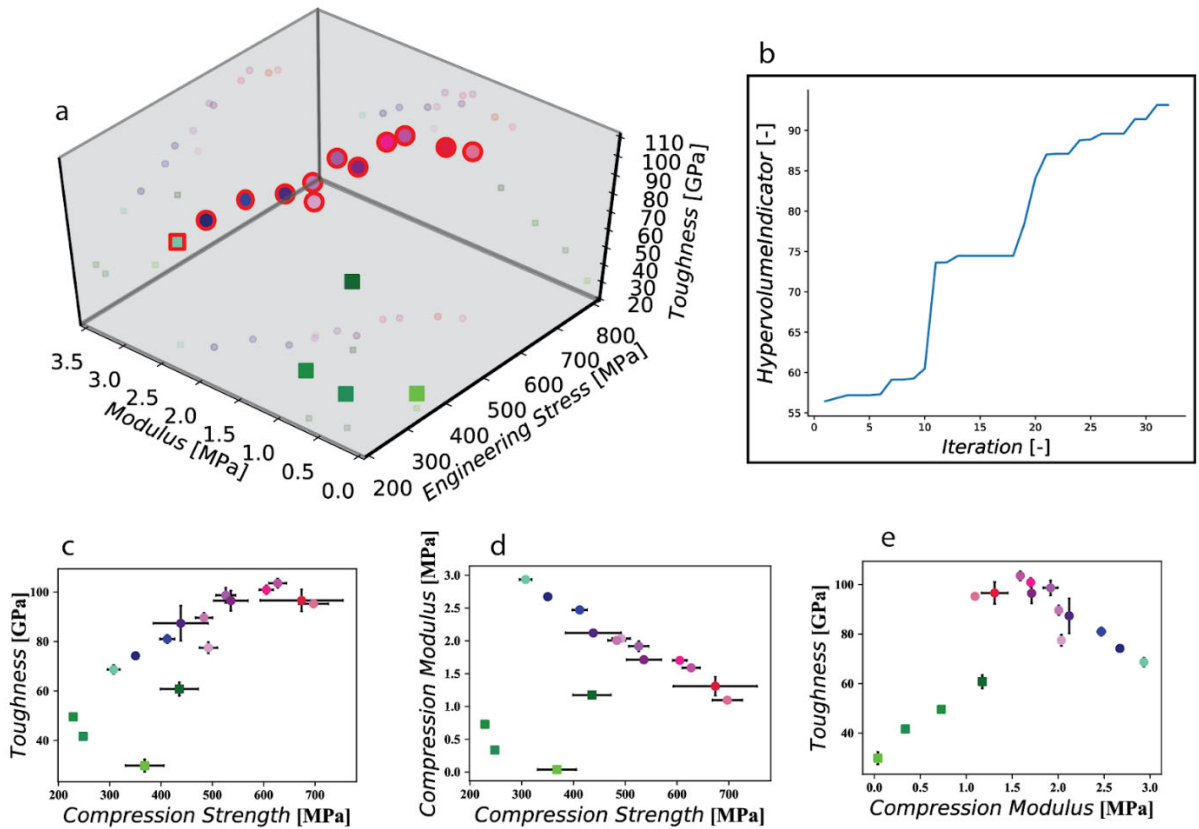

**Fig. 4. Performance results.** (a) 3D plot of the performance space, including the performance of 5 primary formulations A, B, D, E, F in green squares, and the points on the Pareto front obtained after 30 algorithm iterations, circled in red. Primary C is missing as it is too brittle to test in a pure form. Note that primary ink F lies on the Pareto front. (b) Hypervolume improvement plot. Hypervolume indicator shows actual improvement of the Pareto front over iterations of the optimization. (c)-(e) Corresponding 2D plots of pairs of performance objectives.

Aside from discovering the set of optimal solutions, our algorithm that guides the design search also rapidly expands the span of the discovered performance space (see **Figure 5**). The algorithm encourages exploration of unknown regions of the performance space and finds materials with a larger variability in properties. When monitoring the compression strength and compression modulus performance of primary formulations and all evaluated samples, the performance space is enlarged by 250%. The enlargement in compression strength and toughness is larger and rises by 399%. In the view of compression modulus and toughness, the performance space is improved by 584%. This improvement could be important for applications where a specific property range is required and cannot be easily discovered manually.

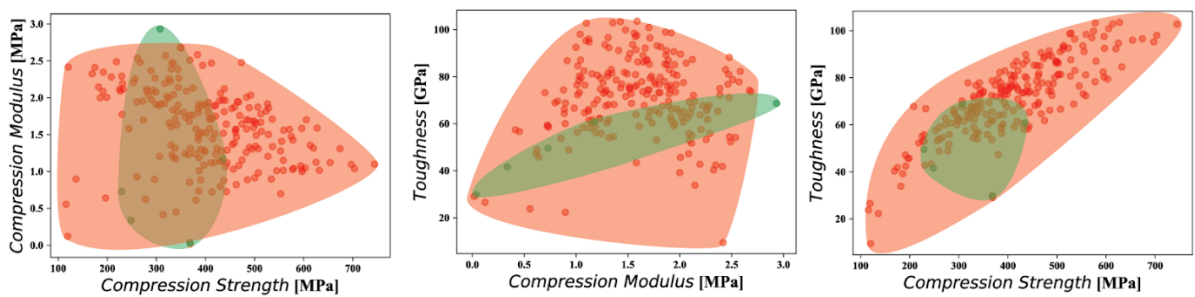

**Fig. 5. Performance space coverage.** For each pair of performance objectives, initial coverage of the performance space is shown in green (spanned by the primary formulations) and the coverage after 30 algorithm iterations is shown in red.

**Discussion**

The discovered formulations in the Pareto set are useful materials in the design of compliant structures where material with an application tailored modulus must be selected to attain the desired outcome[31][32]. For compliant and metamaterial mechanisms that require an actuation force can be tailored by changing the modulus of the materials. To increase the required actuation force, higher modulus materials can be used. Tunable toughness properties are also important for aerospace, packaging and medical applications to reduce mechanical failure and ensure design usability.

The dataset produced by the optimization also provides interesting observations about the chemical compositions, discussed below and graphically presented in Figure 6. Urethane dimethacrylate (UDMA), the main component in base mixture F, is seen to have a large contribution to materials with a high modulus. This contribution is likely due to its high conversion rate and tendency to form hydrogen bonds[29]. Additionally, we see the tendency of the optimization engine to minimize the contribution of hexafunctional aliphatic urethane acrylate, a highly crosslink reagent that leans towards brittle prints. High toughness performance is attained by having formulations with oligomers of urethane-modified acrylate in the amount of 24% to 37%, aliphatic urethane di-acrylate in the amount up to 26%, and UDMA in the amount up to 40%. High toughness formulations also have diluents acrylic acid amide and acrylic acid ester in the range of 14% to 18% and 1% to 19%, respectively. The highest performing compression strength composite formulations included oligomers urethane-modified acrylate at 34%, aliphatic urethane di-arylate at 26%, and UDMA at 6%. They also include diluents acrylic acid amide at 15% and acrylic acid ester at 19%.

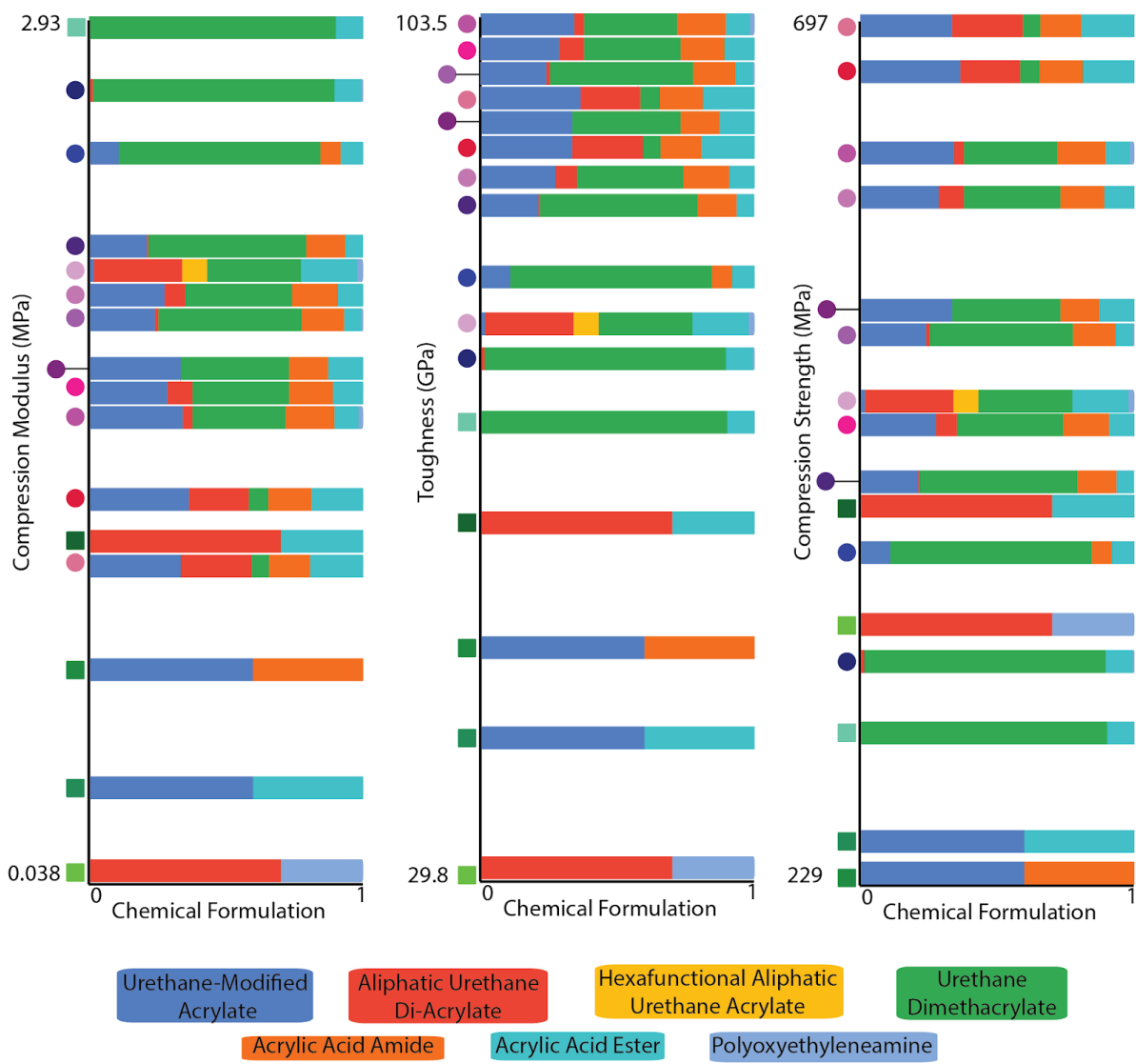

**Fig. 6. Pareto optimal compositions.** Comparison of Pareto optimal (circle) performance characteristics with initial primary formulations (green squares) shows the relationship between performance and formulation ingredients. Base formulation ingredients compositions are indicated via the bar graphs in the above chart. Note that the highest compression modulus point is both the primary formulation point (green square) and Pareto optimal.

The system presented in this paper provides an automation-ready pipeline for improving the performance characteristics of a mixed polymer system. The process is designed for automation with the initial base compositions having a much lower viscosity range to handle than starting from pure polymers. Additionally, automated equipment exists at all steps of the pipeline, from mixing to sample machining, every step of the process can be fully automated. This provides a template for a process that can be adapted to various optimizations, such as coatings or molding, via alteration of the base materials used in the experiments.

While this study presents many advances in multi-objective optimization for 3D printing, it is not without limitation. When defining the design space, base compositions were limited to human chosen inks that were known to print. This provides an improvement in efficiency as all possible combinations are printable, however it may miss some combinations that lie outside of the combination of base inks. Similarly, an improvement could be made to the initial choosing of base compositions via an algorithmic selection mechanic, or an expansion to compose using individual chemicals instead of base compositions. The selection of jet-valve dispensing as the printing process allows for a large range of materials to be considered, however prevents the direct application of the results to commercial printing processes. The same study could be performed on a commercial printing platform to develop application-ready inks.

In summary, the material discovery system described here provides a new way of optimizing photopolymer formulations for additive manufacturing. Utilizing this system allowed us to discover a set of 3D printing material formulations with optimal trade-offs in the mechanical properties of compression modulus, compression strength and toughness. Although this study focuses on the optimization formulations for 3D printing photopolymers, this system can be applied to optimizing materials for other performance objectives, as well as optimizing materials for other manufacturing processes. Our material discovery system lays a foundation for a new tool for material engineers and polymer chemists to discover and optimize materials formulations for a variety of performance objectives and applications.

**Materials and Methods**

**Materials**

Modified Urethane Di-Acrylate, Aliphatic Urethane Di-Acrylate, and Polyoxyethyleneamine and Bis(2,4,6-trimethylbenzoyl)-phenylphosphineoxide are provided by BASF. Aliphatic Urethane Hexa-Acrylate is acquired from Sartometer. Acrylic Acid Amide and Acrylic Acid Ester are provided by Rahn AG. Tween-20 (Polysorbate 20) and Urethane Dimethacrylate are purchased from Sigma-Aldrich.

**Making 3D printable primary inks**

Both viscosity and surface tension limit the compatibility of potential formulations with mixing equipment. Using a jet-valve based 3D printing technology also encounters limitations on both viscosity and surface tension. Lower limitations are imposed by the tendency of the ink to leak from the valve, or spread significantly upon contacting the print platform yielding low print quality. An upper bound on these properties is formed by the inability of the ink to properly separate from the printhead, preventing droplets from forming and causing the printhead to clog. Studies are performed to ensure that the rheology of the inks remained stable while mixing across the design space.

**Formulation Inks Compositions**

Six primary formulations are composed to form as the basis for the design space. The primary formulations consist of four main components, oligomer, reactive diluent, photoinitiator and surfactant. The primary formulations A-F are given in Table S2. The oligomer is the main component of each primary ink, transferring the majority of physical properties to the final printed sample. The reactive diluent serves to reduce the viscosity of the ink to bring it to a printable level. The diluent also factors into the final material properties, as it becomes integrated into the final material structure. The photoinitiator component turns UV light into free radicals to polymerize the material and crosslink the oligomer and reactive diluent together. Surfactant components are added to reduce the surface tension which aids in droplet formation during the printing process.

**Handling and Experimental Cost**

The sample fabrication pipeline operates in a semi-automated manor, where many of the steps can be completed without human input. The dispensing, mixing, 3D printing, post-processing, and testing steps can be completed without human input. Labor is needed to transfer materials between different steps of the sample fabrication pipeline.

The advantage of utilizing the sample fabrication pipeline is to produce and test formulations at a minimal cost. To do this the sample fabrication pipeline design has been optimized. The duration of each step of the sample fabrication pipeline is listed in Table S1.

**Dispensing**

Dispensing of the primary formulation is done using a custom syringe pump-based system. The system was constructed using stepper motors driving linear rails to compress 100 mL glass syringes. Syringe outputs are all routed to a manifold which directs the primary formulations into a single mixing cup. The system is controlled using a microcontroller (Arduino Mega 2560, Turin, Italy) with 6-axis stepper control software (GBRL, opensource software). This software synchronizes all of the movement commands, so primary formulations can be simultaneously dispensed at varying speeds. To dispense a specific primary formulation, the displacement of each syringe is calculated to be proportionate to the primary formulation's contribution to the composite formulation. This six-axis command is then issued to the controller. Gathering only the steady state flow from the dispenser allows for differences in the surface tension and viscosity of the differing primary formulations to be circumvented.

**Mixing**

Mixing components together takes a significant amount of the total time used for preparing formulations for 3D printing. The mixing process is designed to mitigate the amount of time it takes to mix a formulation during processing. To reduce the mixing time down to a more acceptable range, a SpeedMixer (FlackTek, Landrum, SC, USA) is utilized to mix the components together and is shown in Figure S1. This instrument is a dual axis centrifuge, causing materials within the mixing vessel to flow into themselves, rapidly mixing them together[37,38]. The overall mixing time takes 64.7 seconds at 3000 rpm for each formulation mixed.

**3D Printing**

In order to print 3D printing inks with a wide variety of fluid properties, we designed a hardware setup based on piezo-actuated jet-valve. The jetting head consists of a nozzle with a diameter of 300 microns, a spring-loaded needle valve, a pressurized ink reservoir, a valve body, and a piezoelectric actuator as shown in Figure S2. To create a droplet, first the piezoelectric actuator releases the force holding the needle valve in the closed position. The spring forces the needle valve into its open state. The pressurized ink is forced through the nozzle opening dispensing fluid. Finally, the piezoelectric actuator closes the needle valve into the closed position, cutting off the flow of ink. The amount of formulation that is dispensed is a function of the opening time of the nozzle, the back pressure of the ink, and the ink's fluid properties.

The jet-valve 3D printer is comprised of 2 jet-valve dispensers (PICO Pμlse® from Nordson EFD, Providence, RI, USA), a 365nm UV LED (Phoseon Technology, Hillsboro, OR, USA), a Cartesian robot (Hiwin Mikrosystems, Taichung, Taiwan), a gantry, a build platform and a controller. The jet-valve dispensers are used to dispense two materials in parallel. The UV LED is used to cure the ink after it is dispensed by the dispensers[39]. The Cartesian robot is utilized to change the position of the jet-valve dispensers and UV LED with respect to the build platform. The movement of the Cartesian robot and the actuation of the dispensers and UV LED are controlled by a g-code file that is interpreted by the controller[40].

The controller utilizes the same g-code file for each formulation. To create a layer of the 3D printed sample, the dispenser follows a vector path and deposits droplets of ink to create a continuous line of material. It follows a vector path until an entire layer of material is deposited. After the layer is finished, the UV LED is turned on and rasters over the entire deposited layer to cure the ink. After curing, the dispensers and UV LED are incremented upwards by the thickness of one layer and the process repeats until the prescribed 3D object has been dispensed.

G-code instructions are produced using slicing software (Slic3r, opensource software), which generates printing instructions for a fused deposition modeling process. This code is then post-processed further to insert UV LED steps at the end of layers, and change dispensing instructions to communicate properly with the jet-valve dispensers. Four 8mm by 8mm cylinders in a grid array are imported into the slicing software. Droplet spacing of 0.5 mm was selected to ensure no voids were present between droplets.

While the overall g-code print instructions remained constant for all samples, parameters on the print head controller had to be adjusted to ensure printability and consistency in droplet size across ink formulations. Temperature of the printhead was kept at a steady 60°C for all print, and the close voltage of the valve was also kept at the maximum possible value. Values for syringe pressure, valve opening percentage, and valve opening time were adjusted per print to normalize the mass of an individual droplet. Changing these dispensing parameters result in changes in droplet diameter as shown in Figure S3. As valve opening percentage, syringe pressure and valve opening time increases the size of the droplet increases. These parameters must be over a threshold that ensure a droplet is jetted properly. If jetting parameters are set too high, splashing can occur when the formulation hits the substrate. Jetting parameters were selected manually for each formulation.

**Post Processing of Printed Samples**

To ensure consistency across printed samples with different formulations, we employ a post curing step after printing to complete the polymerization process. This accounts for any reaction rate based differences in the printing process due to differing ink compositions, which could be accounted for in printing process optimization. Typically, one hour of baking at 60ºC and flooding in 365nm/405nm UV light in a post-curing oven (Wicked Engineering, East Windsor, CT, USA) is employed to complete polymerization. Further, to remove variations in the height of the compression samples caused by printing and post-processing [Katheng2020], the samples are flattened with a machining process.

**Mechanical Performance Testing of Printed Samples**

Physical characteristics of maximum stress, maximum toughness and modulus are chosen for the performance space. ASTMD-638 standard was initially chosen for measuring these properties, using the smallest possible dogbone (standard V) allowed in the documentation. This provided for a dogbone of 63.5 mm by 9.5 mm, and an overall volume of approximately 1580 mm$^3$. While useful for initial testing and verification of our method against material data, these samples require precise positioning in an universal testing system (Instron 68SC-1) taking approximately 3 minutes per replicate, and 12 minutes per sample to complete testing.

An alternative testing methodology is desired, to increase throughput and reduce material requirements. Compression testing[41] circumvents the precision positioning step required in tensile testing, reducing testing time. Additionally, sample size has been shrunk to 8 mm in diameter, with an overall volume of approximately 400 mm$^3$. This allows for more samples to be printed on each print platform, reducing printing time and significantly reducing the amount of raw materials to complete optimization.

Compression testing is completed on Instron 5984 Universal Tester. Multiple compression rates were tested on these instruments, finding a maximum compression rate of 3 mm/minute while maintaining consistency in testing data. In addition to optimizing the rate of compression, a machining step was added to the process to remove inconsistencies in the top of the printed samples. These inconsistencies arise from differences in surface tensions between inks, resulting in differing degrees of doming between prints. By roughly machining the printed samples while still attached to the print platform, consistent surfaces on the top and bottom can be achieved in an automation friendly format in under 2 minutes per composition. These samples are then measured in both diameter and height via a caliper to account for dimensional differences. All the samples are measured for 3 times to obtain a reliable result. Mechanical properties are then computed from the stress-strain curve of each sample, averaged, and inputted to the optimization algorithm.

**Performance Data Analysis**

To complement the highly automated process of manufacturing samples, an automated analysis pipeline is also introduced, producing performance parameters from raw stress strain data as shown in Figure S4. Initial data was first transformed into engineering stress by dividing the measured force by the surface area of the sample. Engineering strain was normalized by dividing the displacement of the tester by the original height of the sample. Once converted to engineering values, the stress strain data is then trimmed, allowing for excess data in the initial loading cycle of the tester to be removed along with data after the failure point of the sample. The initial loading segment is trimmed via careful monitoring of the slope of the stress strain data, looking for the slope to increase to at least 0.1, all data prior to this point is discarded as noise in initial loading. To find the failure point, the second derivative of the engineering data was rank ordered and filtered for points with at least three consecutive negative slopes following. Additional filtering on this list of potential failure points was applied removing points that occurred within the first 30% of the data series. These two methodologies were developed using initial datasets, comparing against how a human mechanical engineer would process the stress strain data.

Once the initial data was preprocessed, transformation into performance parameters could occur. The modulus was determined by calculating a linear fit on the engineering stress strain data for all points below 100 MPa with a sliding window of 20 data points. The median modulus is then calculated from these values and a filter for all moduli within 60% of the median is applied. Using the values within 60% deviation from the median, a final linear fit is applied across the entire range and classified as the modulus for the sample. Similar to the trimming of the stress strain data, these calculated values are compared to human calculated values from a mechanical engineer for the initial dataset. The maximal values for both engineering stress and strain are taken from the trimmed data and used for performance values.

**Design Space and Initial Samples**

A design space is a 6-dimensional space, where 5 parameters can take real values from 0 to 1, and 1 parameter is constrained from 0 to 0.5. To simplify the fabrication, we limit the parameters to values with 2 decimals due to resolution limitations in dispensing. In that case, 5 parameters can take 101 different values, and the sixth parameter can take 51 different values. Furthermore, these parameters need to sum up to 1. All valid combinations of these parameters then lie on a standard 5-simplex, a 5-dimensional polytope that is a convex hull of its 6 vertices.

The set of 25 initial samples is generated to uniformly cover the design. We get uniformly distributed points on the standard simplex by generating a random vector of 5 values from the symmetric Dirichlet distribution.

**Multi-Objective Optimization Algorithm**

We start by evaluating 25 randomly generated samples that try to cover the design space as uniformly as possible and 5 primary formulations, having 30 initial samples in total. The first part of the algorithm fits a Gaussian Process (GP) for each objective independently. The GPs are trained on the evaluated data points and are used as a surrogate model for black-box objective functions. The Thompson sampling of the GPs is then used to balance the trade-offs of exploiting the best performing regions and exploring unseen regions of the design space[28]. The third part of the algorithm solves a multi-objective optimization problem on the previously sampled functions. This part extracts a predicted Pareto front and Pareto set. The final stage of the algorithm proposes which samples to evaluate in the next iteration. To further reduce the experimental time, we propose a batch of 4 samples to be evaluated in parallel. Four samples with the largest expected hypervolume improvement from the predicted Pareto front are chosen. We then evaluate the proposed samples, add them to the currently available dataset and iterate through the algorithm.

**Algorithm Parameters**

More details on each part of the algorithm and general hyperparameters can be found in Bradford et al.[18]. Here, we list hyperparameters and algorithm alterations used in our implementation.

In each iteration of the algorithm, a Gaussian process (GP) is fitted for each objective independently. To train the hyperparameters of GP, the maximum a posteriori estimate is used, as proposed in Bradford et al.[18]. We use a Matérn kernel 5/2, as it can support the most general and complex function types.

To solve the multi-objective optimization problem on objective functions extracted with Thompson sampling from GPs, a standard NSGA-II solver[36] is used. In each iteration, we use a population size of 100, and a total number of 100 generations. Handling the linear equality constraint $a+b+c+d+e+f = 1$ is done by adding two inequality constraints to the solver, $a+b+c+d+e+f -1 \leq 0$ and $a+b+c+d+e+f -1 \geq 0$, making sure that mutation points that do not satisfy these constraints are not proposed.

To monitor the hypervolume improvement over iterations, we use a fixed reference point [1.361656338114889e+02, 0.037819494001910, 2.226856385006103e+04] throughout the performance space. This point is chosen because it has a minimal value for each of the 3 objectives of evaluated points in the initial dataset.

**References**

[1] R. L. Truby, J. A. Lewis, Printing Soft Matter in Three Dimensions. Nature 540, 371-378 (2016).

[2] S. C. Ligon, R. Liska, J. Stampfl, M. Gurr, R. Mülhaupt, Polymers for 3D Printing and Customized Additive Manufacturing. Chem. Rev. 117, 10212 (2017).

[3] N. W. Bartlett, M. T. Tolley, J. T. Overvelde, J. C. Weaver, B. Mosadegh, K. Bertoldi, G. M. Whitesides, R. J. Woods, A 3D-printed, functionally graded soft robot powered by combustion. Science. 349, 161 (2015).

[4] J. Ten Kate, G. Smit, P. Breedveld, 3D-printed upper limb prostheses: a review. Disabil. Rehabil.: Assist. Technol. 12, 300 (2017).

[5] A. V. Do, B. Khorsand, S. M. Geary, A. K. Salem 4, 3D printing of Scaffolds for tissue regeneration applications. Adv. Healthcare Mater. 4, 1742 (2015).

[6] J. Tumbleston, D. Shirvanyants, N. Ermoshkin, R. Janusziewicz, A. Johnson, D. Kelly, K. Chen, R. Pinschmidt, J. Rolland, A. Ermoshkin, E. Samulski, J. DeSimone, Continuous liquid interface production of 3D objects. Science. 347, 1349 (2015).

[7] F. Kotz, K. Arnold, W. Bauer, D. Schild, N. Keller, K. Sachsenheimer, T. M. Nargang, C. Richter, Three-dimensional printing of transparent fused silica glass. Nature. 544, 337 (2017).

[8] K. Sun, T. S. Wei, B. Y. Ahn, J. Y. Seo, S. J. Dillon, J. A. Lewis, 3D printing of interdigitated Li-Ion microbattery architectures. Adv. Mater. 25, 4539 (2013).

[9] T. S. Wei, B. Y. Anh, J. Grotto, J. A. Lewis, 3D printing of customized li-ion batteries with thick electrodes. Adv. Mater. 30, 1703027 (2018).

[10] D. W. McOwen, S. Xu, Y. Gong, Y. Wen, G. L. Godbey, J. E. Gritton, T. R. Hamann, J. Dai, G. T. Hitz, L. Hu, E. D. Wachsman, 3D-printing electrolytes for solid-state batteries. Adv. Mater. 30, 1707132 (2018).

[11] C. Wang, W. Ping, Q. Bai, H. Cui, R. Hensleigh, R. Wang, A. H. Brozena, Z. Xu, J. Dai, Y. Pei, C. Zheng, A general method to synthesize and sinter bulk ceramics in seconds. Science. 368, 521 (2020).

[12] R. Epps, M. Bowen, A. Volk, K. Abdel-Latif, S. Han, K. Reyes, A. Amassian, M. Abolhasani, Artificial Chemist: An Autonomous Quantum Dot Synthesis Bot. Adv. Mater. 32, 2001626 (2020).

[13] C. W. Coley, D. A. Thomas, J. A. Lummiss, J. N. Jaworski, C. P. Breen, V. Schultz, T. Hart, J. S. Fishman, L. Rogers, H. Gao, R. W. Hicklin, A robotic platform for flow synthesis of organic compounds informed by AI planning. Science. 365, 557 (2019).

[14] A. C. Bédard, A. Adamo, K. C. Aroh, M. G. Russell, A. A. Bedermann, J. Torosian, B. Yue, K. F. Jensen, T. F. Jamison, Reconfigurable system for automated optimization of diverse chemical reactions. Science. 361, 1220 (2018).

[15] B. Grigoryan, S. J Paulsen, D. C. Corbett, D. W. Sazer, C. L. Fortin, A. J. Zaita, P. T. Greenfield, N. J. Calafat, J. P. Gounley, A. H. Ta, F. Johansson, Multivascular networks and functional intravascular topologies within biocompatible hydrogels. Science. 364, 458 (2019).

[16] J. Chang, P Nikolaev, J. Carpena-Nunez, R. Rao, K. Decker, A. Islam, J Kim, M. Pitt, J. Myung, B. Maruyama, Efficient Closed-loop Maximization of Carbon Nanotube Growth Rate using Bayesian Optimization. Sci. Rep. 10, 9040 (2020).

[17] B. P. Macleod, F. G. Parlane, T. D. Morrissey, F. Häse, L. M. Roch, K. E. Dettelbach, C. P. Berlinguette, Self-driving laboratory for accelerated discovery of thin-film materials. Sci. Adv. 6, EAAZ8867 (2020).

[18] E. Bradford, A. Schweidtmann, A. Lapkin, Efficient multiobjective otpimization employing Gaussian processes, spectral sampling and a genetic algorithm. J. Glob. Optim. 71, 407–438 (2018).

[19] U. G. Wegst, H. Bai, E. Saiz, A. P. Tomsia, R. O. Ritchie, Bioinspired structural materials. Nat. Mater. 14, 23 (2015).

[20] B. Burger, P. M. Maffettone, V. V. Gusev, C. M. Aitchison, Y. Bai, X. Wang, X. Li, B. M. Alston, B. Li, R. Clowes, N. Rankin, B. Harris, R. S. Sprick, A. I. Cooper, A mobile robotic chemist. Nature 583, 237–241 (2020).

[21] A. E. Gongora, B. Xu, W. Perry, C. Okoye, P. Riley, K. G. Reyes, E. F. Morgan, K. A. Brown, A Bayesian experimental autonomous researcher for mechanical design. Sci. Adv. 6, p.eaaz1708 (2020).

[22] M. Layani, X. Wang, S. Magdassi, Novel materials for 3D printing by photopolymerization. Adv. Mater. 30, 1706344 (2018).

[23] J. Zhang, P. Xiao, 3D printing of photopolymers. Polym. Chem. 9, 1530 (2018).

[24] H. Li, J. Liu, K. Li, Y. Liu, Piezoelectric micro-jet devices: A review. Sens. Actuators, A. 297, 111552 (2019).

[25] H. Yin, Y. Ding, Y. Zhai, W. Tan, X. Yin, Orthogonal programming of heterogeneous micro-mechano-environments and geometries in three-dimensional bio-stereolithography. Nat. Comm. 9, 4096 (2018).

[26] B. Shahriari, K. Swersky, Z. Wang, R. P. Adams, and N. de Freitas, Taking the human out of the loop: A review of Bayesian optimization. Proceedings of the IEEE. 104, 148–175 (2016).

[27] P. E. Antonio, D. F. González, and L. F. Verdeja, Selection of Structural Materials: Combined Mechanical Properties and Materials-Selection Charts. Structural materials. Springer, Cham, Switzerland. (2019).

[28] B. Shahriari, Z. Wang, M.W. Hoffman, A. Bouchard-Côté, N.D. Freitas, An entropy search portfolio for Bayesian optimization. ArXiv. abs/1406.4625 (2014).

[29] I. Barszczewska-Rybarek, A Guide through the Dental Dimethacrylate Polymer Network Structural Characterization and Interpretation of Physico-Mechanical Properties. Materials. 12(24), 4057 (2019).

[30] E. Yang, M. Leary, B. Lozanovski, D. Downing, M. Mazur, A. Sarker, A. Khorasani, A. Jones, T. Maconachie, S. Bateman, M. Easton, M. Qian, P. Choong, M. Brandt. Effect of geometry on the mechanical properties of Ti-6Al-4V Gyroid structures fabricated via SLM: A numerical study. Materials & Design. 184, 108165 (2019).

[31] J. Lipton, R. Maccurdy, Z. Manchester, L. Chin, D. Cellucci, D. Rus, Handedness in shearing auxetics creates rigid and compliant structures. Science, 360(6389), 632-635 (2018).

[32] A. Ion, J. Frohnhofen, L. Wall, R. Kovacs, M. Alistar, J. Lindsay, P. Lopes, H. Chen, P. Baudisch, Metamaterial Mechanisms. Proceedings of the 29th Annual Symposium on User Interface Software and Technology, 529-539 (2016).

[33] J. Sun, X. Zhao, W. Illeperuma, O. Chaudhuri, K. Hwan Oh, D. Mooney, J. Vlassak, Z. Suo, Highly stretchable and tough hydrogels. Nature, 489, 133-136 (2012).

[34] N. Annabi, Y. Zhang, A. Assmann, E. Sani, G. Cheng, A. Lassaletta, A. Vegh, B. Dehghani, G. Ruiz-Esparza, X. Wang, S. Gangadharan, A. Weiss, A. Khademhosseini, Engineering a highly elastic human protein–based sealant for surgical applications. Sci. Transl. Med., 9, 410 (2017).

[35] M. Möller, D. Kunz, T. Lunkenbein, S. Sommer, A. Nennemann, J. Breu, UV-Cured, Flexible, and Transparent Nanocomposite Coating with Remarkable Oxygen Barrier. Adv. Mater, 24(16), 2142–2147 (2012).

[36] K. Deb, A. Pratap, S. Agarwal, T. Meyarivan, A fast and elitist multiobjective genetic algorithm: NSGA-II. IEEE Transactions on Evolutionary Computation, 6(2), 182-197 (2002).

[37] P. Darwin, G. Seeley, M. Gio-Batta, A. Burgess, Structure and properties of epoxy-based layered silicate nanocomposites. J. Macromol. Sci., 44(6), 1021-1040 (2005).

[38] G. Wang, X. Chen, R. Huang, L. Zhang, Nano-caco3/polypropylene composites made with ultra-high-speed mixer. J. Mater. Sci. Lett., 21(13), 985-986 (2002).

[39] Y. Kim, S. Hong, H. Sun, M. Kim, K. Choi, J. Cho, H. Choi, J. Koo, H. Moon, D. Byun, K. Kim, J. Suhr, S. Kim, J. Nam, UV-curing kinetics and performance development of in situ curable 3D printing materials. Eur. Polym. J., 93, 140-147 (2017).

[40] A. Foerster, R. Wildman, R. Hague, C. Tuck, Reactive inkjet printing approach towards 3D silicone elastomeric structures fabrication. Solid Freeform Fabrication Symposium (2017).

[41] A. Detwiler, A. Lesser, Characterization of double network epoxies with tunable compositions. J. Mater. Sci., 47, 3493-3503 (2012).

**Acknowledgments**

**General**: We would like to thank Prof. Alan Lesser, Chimney Saraf, and Brendan Ondra for very helpful discussions about a rational mechanical testing strategy at the beginning of this work. M. K. Luković would like to acknowledge support from the Schmidt Science Fellowship.

**Funding:** The research was supported by BASF.

**Author contributions:** T. E. and M. F. contributed equally to this work, working on the physical construction of the system, manufacture and testing of samples. M. K. L. contributed to development of the optimization strategy and algorithm implementation used in this paper. W. S., H. G., H. D., K. S., B vV. and W. M. provided invaluable feedback throughout the course of the project, helping to guide the path of the investigation.

**Competing interests:** The authors declare no competing interests.

**Data and materials availability:** All data needed to evaluate the conclusions in the paper are present in the paper and/or the Supplementary Materials.

# Supplementary Materials

**Table S1.** The average duration and standard deviation of each time processing step within the sample fabrication pipeline.

|  | Dispensing | Mixing | Syringe Loading | 3D Printing | Post-Processing | Testing | **Total** |
|---|---|---|---|---|---|---|---|
| Average Time | 157.7 s | 64.7 s | 45.7 s | 1975 s | 4354 s | 199.7 s | **6797 s** |
| Standard Deviation | 3 s | 0.6 s | 3.2 s | 190.9 s | 13.54 s | 30 s | **241 s** |

**Table S2:** Formulations of the 6 primary inks (weight percentage).

| Primary Formulation | Oligomer | Diluent | Photoinitiator | Surfactant |
|---|---|---|---|---|
| A | 59% Modified Urethane Di-Acrylate | 39.5% Acrylic Acid Amide | 0.5% Bis(2,4,6-trimethylbenzoyl)-phenylphosphineoxide | 1% Tween-20 |
| B | 69% Aliphatic Urethane Di-Acrylate | 30% Acrylic Acid Ester | 0.5% Bis(2,4,6-trimethylbenzoyl)-phenylphosphineoxide | 0.5% Tween-20 |
| C | 69% Aliphatic Urethane Hexa-Acrylate | 30% Acrylic Acid Ester | 0.5% Bis(2,4,6-trimethylbenzoyl)-phenylphosphineoxide | 0.5% Tween-20 |
| D | 69% Aliphatic Urethane Di-Acrylate | 30% Polyoxyethyleneamine | 0.5% Bis(2,4,6-trimethylbenzoyl)-phenylphosphineoxide | 0.5% Tween-20 |
| E | 59% Modified Urethane Di-Acrylate | 40% Acrylic Acid EsterRahn Genomer 1121 | 0.5% Bis(2,4,6-trimethylbenzoyl)-phenylphosphineoxide | 0.5% Tween-20 |
| F | 89.5% Urethane Dimethacrylate | 10% Acrylic Acid Ester | 0.5% Bis(2,4,6-trimethylbenzoyl)-phenylphosphineoxide |  |

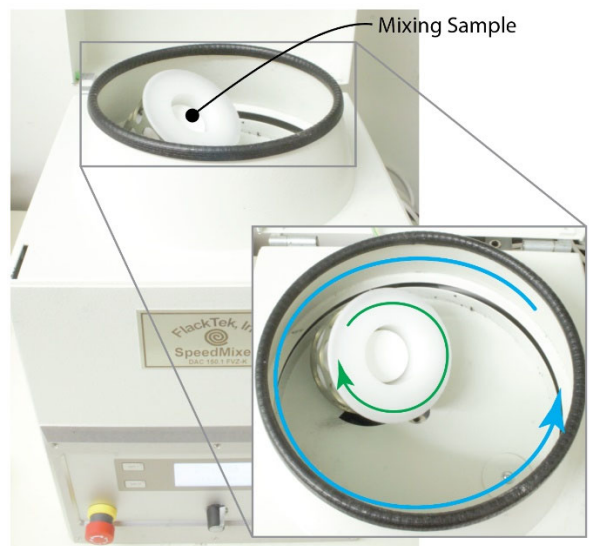

**Fig. S1. Formulation mixer.** Dual asymmetric centrifuge mixer utilized for sample fabrication pipeline. The arrows depict the rotation of the sample during operation.

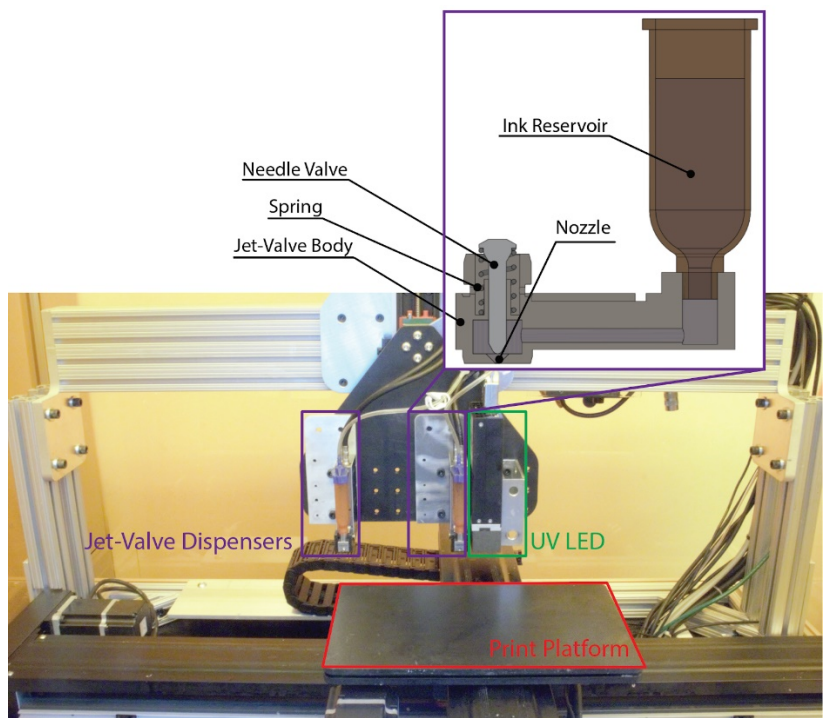

**Fig. S2. Jet-valve 3D printer utilized in the sample fabrication pipeline.** Circled in purple are the 2 jet-valve dispensers used for dispensing the material in parallel. Circled in green is the UV LED used for curing the dispensed material. Circled in red is the print platform, the area where the samples are dispensed onto and cured.

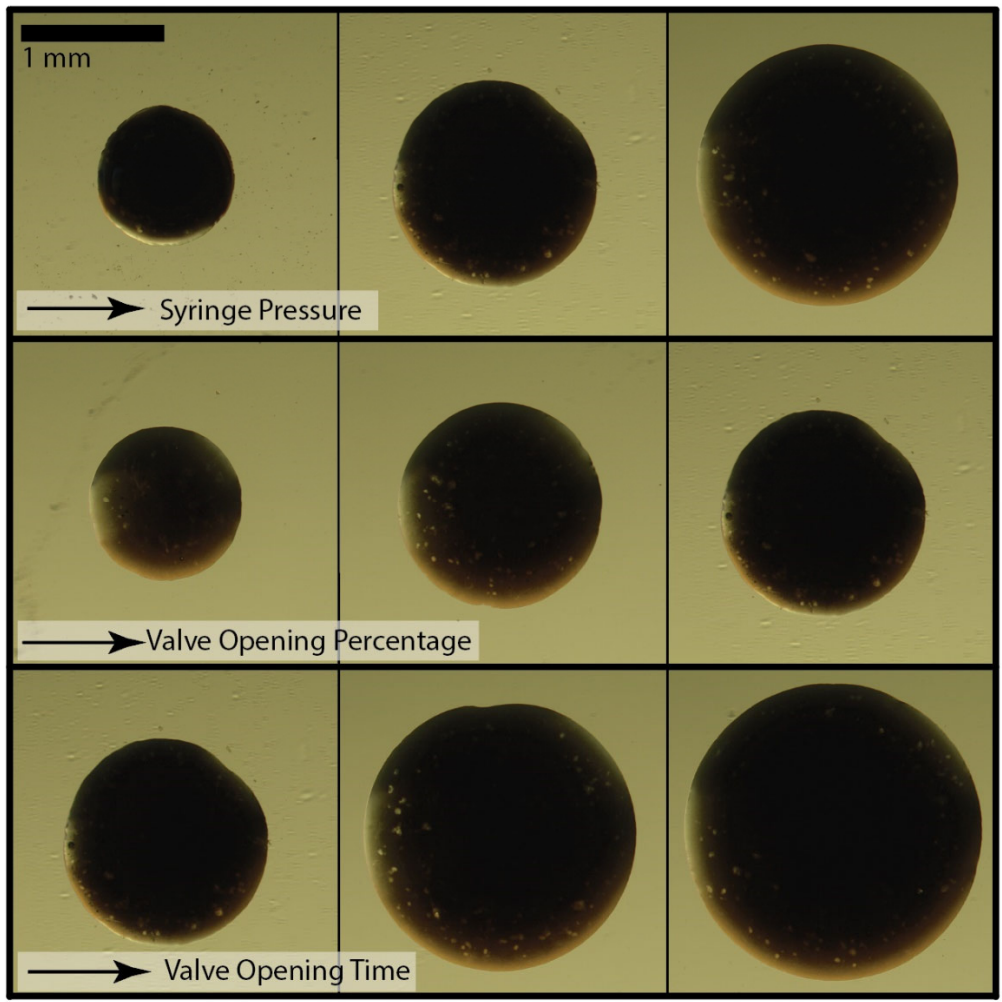

**Fig. S3. Optical Microscopy images of jet-valve dispensing droplets.** Droplet variations in Syringe pressure (top) (parameters from left to right: 15 psi, 35 psi, 50 psi, all 90 % opening percentage and 1 ms opening time), valve opening pressure (middle) (parameters from left to right: 70 %, 80 %, 90 %, all 35 psi syringe pressure and 1 ms opening time), and valve opening time (bottom) (parameters from left to right: 1 ms, 1.5 ms, 2 ms, all 35 psi syringe pressure and 90 % opening percentage).

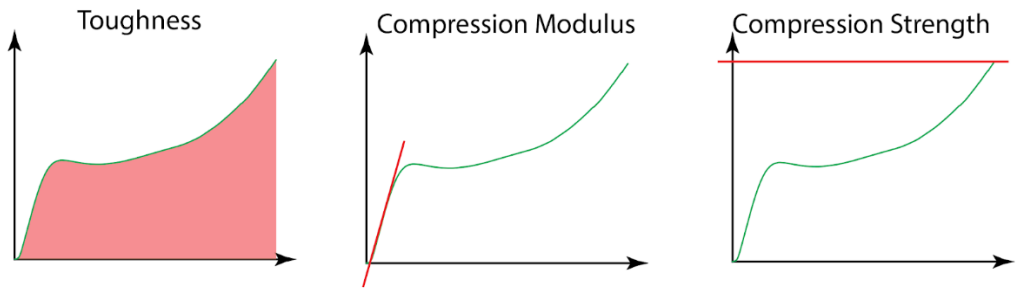

**Fig. S4. The 3 performance objectives that formulations are optimized for.** Performance objectives are parsed from stress-strain data that is resultant of the compression testing.

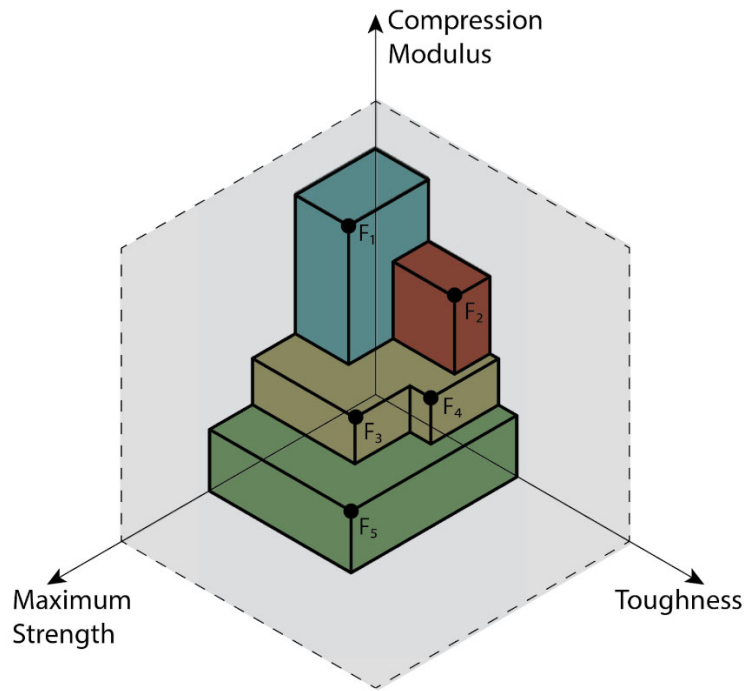

**Fig. S5. Hypervolume calculation.** A hypervolume that is calculated on Pareto optimal points $F_1$, $F_2$, $F_3$, $F_4$, and $F_5$, with an origin as a reference point.